\documentclass[conference]{IEEEtran}
\ifCLASSINFOpdf
\else
\fi

\usepackage{hyperref}
\usepackage{xcolor}
\usepackage{graphicx}
\usepackage{listings}
\usepackage{comment}
\usepackage{mdframed}
\definecolor{dkgreen}{rgb}{0,0.6,0}
\definecolor{darkred}{rgb}{0.3,0.1,0.1}
\definecolor{gray}{rgb}{0.5,0.5,0.5}
\definecolor{mauve}{rgb}{0.58,0,0.82}
\definecolor{mGray}{rgb}{0.5,0.5,0.5}
\definecolor{mPurple}{rgb}{0.58,0,0.82}

\newcommand{\TASK}[1]{
  {\begin{mdframed}[linecolor=mPurple]
  \noindent\textcolor{mPurple}{\textbf{TASK:} \em\footnotesize#1}
  \end{mdframed}}
  {}
}

\newcommand{\TODO}[1]{
  {\begin{mdframed}[linecolor=red]
  \noindent\textcolor{red}{\textbf{TODO:} \em\footnotesize#1}
  \end{mdframed}}
  {}
}

\newcommand{\standard}{{Standard}}
\newcommand{\mpi}{{MPI}}

\lstdefinestyle{DerekStyle}{
  language=C++,
  basicstyle=\small\ttfamily,
  numbers=left,
  numberstyle=\scriptsize\color{mGray},  
  numbersep=5pt,                  
  backgroundcolor=\color{white},
  showspaces=false,               
  showstringspaces=false,         
  showtabs=false,
  frame=single,                   
  rulecolor=\color{black},
  tabsize=2,                      
  captionpos=b,                   
  breaklines=true,                
  breakatwhitespace=false,
  keywordstyle=\color{blue},          
  commentstyle=\color{dkgreen},       
  stringstyle=\color{mauve},         
  escapeinside={\%*}{*)},
  xleftmargin=4.0ex,
  morekeywords={for,each,between,can,reach,in,is,Sort,Print,From,MPI_Count,MPI_Comm}
}

\usepackage{todonotes} 

\begin{document}
%
\title{MPI's Language Bindings are Holding MPI Back}



\author{
\IEEEauthorblockN{Martin Ruefenacht}
\IEEEauthorblockA{Leibniz Supercomputing Centre\\
Bavaria, Germany\\
\textit{\href{mailto:martin.ruefenacht@lrz.de}{martin.ruefenacht@lrz.de}}}
\and
\IEEEauthorblockN{Derek Schafer and Anthony Skjellum}
\IEEEauthorblockA{University of Tennessee at Chattanooga\\
Chattanooga, Tennessee, USA \\
\textit{\href{mailto:derek-schafer@utc.edu}{derek-schafer@utc.edu}, \href{mailto:Tony-Skjellum@utc.edu}{tony-skjellum@utc.edu}}}
\and
\IEEEauthorblockN{Purushotham V. Bangalore}
\IEEEauthorblockA{University of Alabama at Birmingham \\ Birmingham, Alabama, USA\\ \textit{\href{mailto:puri@uab.edu}{puri@uab.edu}}}
}


%


\maketitle

\begin{abstract}
Over the past two decades, C++ has been adopted as a major HPC language (displacing C to a large extent, and Fortran to some degree as well). Idiomatic C++ is clearly how C++ is being used nowadays. But, MPI's syntax and semantics defined and extended with C and Fortran interfaces that align with the capabilities and limitations of C89 and Fortran-77. Unfortunately,  the  language-independent specification also clearly reflects the intersection of what these languages could syntactically and semantically manage at the outset in 1993, rather than being truly language neutral.

In this paper, we propose a modern C++ language interface to replace the C  language binding for C++ programmers with an upward-compatible architecture that leverages all the benefits of C++11--20 for performance, productivity, and interoperability with other popular C++ libraries and interfaces for HPC.
Demand is demonstrably strong for this second attempt at language support for C++ in MPI after the original interface, which was added in MPI-2, then was found to lack specific benefits over the C binding, and so was subsequently removed in MPI-3.

Since C++ and its idiomatic usage have evolved  since the original C++ language binding was removed from the standard, this new effort is both timely and important for MPI applications.  Also, many C++  application programmers create their own, ad hoc shim libraries over MPI to provide some degree of abstraction unique to their particular project, which means many such abstraction libraries are being devised without any specific commonality other than the demand for such.

\end{abstract}


%
\IEEEpeerreviewmaketitle


\section{Introduction}\label{sec:intro}

 The Message Passing Interface (\mpi{}) has been the ubiquitous programming model for scalable computing since 1994.  But, as \mpi{} ages, it is  clear it is held back in terms of function and performance by the close ties between the concepts that constitute \mpi{} and the languages in which it was originally expressed. This drawback was most recently evidenced by the introduction of the large-count procedures, which doubled 157 of the original \mpi{} 3.1 procedures~\cite{BigMPI,Embiggenment,MPI-3.1}. In  other situations, potential features of \mpi{} are not broached as topics of discussion because they are not easily expressible in either C or Fortran, or the people who would be interested in such a feature do not use \mpi{} since they do not use either C or Fortran. Overall, \mpi{} has  been  slow to adapt and/or embrace  new    programming trends and complementary technologies.

The \mpi{} \standard{}, which provides C and Fortran as official languages, has been unofficially augmented with community-supported  language bindings. Some languages that are supported include C\#, Go, Julia, Python, R, Rust, among many others. From this, we surmise that demand exists for support for the \mpi{} concepts in many languages and that the concepts are appropriate to be used in many situations. In addition to the community-supported bindings, large C++ applications have also implemented internal bindings for \mpi{}, which map the C interface to their preferred internal C++ interface (e.g., ~\cite{COMB}).


The gap addressed by these community-led projects is that \mpi{}, as it is currently standardized, does not allow for modern applications or languages to make use of it idiomatically. The language-independent specification effectively gates  \mpi{} concepts behind the C89 and Fortran-77 languages. 

Here, 
we propose the official introduction of higher level languages to the \mpi{} \standard{} in addition to the standardized C and Fortran interfaces. By introducing a common layer for a given language, we fulfill the same goal as the originally standardized languages: to enable development of an \mpi{} application with performance-portability between implementations. For each such standardized language, we could choose to accommodate language features that are either difficult or impossible to support effectively through a community-led project based on the current \mpi{} \standard{}. This approach enables expression of \mpi{} idiomatically for each language while keeping the core concepts uniform across all expressed languages. First-class \mpi{} concepts are thereby standardized for all languages while each language expresses those semantics and operations naturally.
%
We use C++ as the target language in the bulk of this paper, and show how even minimal use of C++ can make huge improvements over the C interface.


The remainder of this paper is organized with  Section~\ref{sec:mpicpp} discussing the \mpi{} C++ interface that was introduced in \mpi{} 2.0, and mention its shortcomings; then, Section~\ref{sec:badc} examines how the C \mpi{} interface holds  \mpi{} back as a whole. Section~\ref{sec:related} explores approaches that have been taken in the community to introduce higher level language interfaces to \mpi{}. Section~\ref{sec:consequences} considers the consequences of introducing new language interfaces to \mpi{}. Section~\ref{sec:idiomatic} looks at potential C++ features of which MPI could take advantage. Section~\ref{sec:zeroth} introduces a minimally extended C++ language interface to \mpi{} that attempts to remain as close to the present C interface as possible while still introducing C++ concepts. 
Section~\ref{sec:mixing} discusses mixing C++ and C \mpi{} code and  libraries. Finally, we conclude and offer final recommendations in Section~\ref{sec:conclusions}.

\section{The Failure of C++ Bindings in MPI 2.0}\label{sec:mpicpp}

The \mpi{} C++ bindings extended the \mpi{} C object-based design and wrapped the key \mpi{} concepts as classes and for the most part retained a one-to-one correspondence between the C and C++ APIs. This was a deliberate decision of the \mpi{} Forum~\cite{DBLP:conf/iscope/SquyresSL97}; many of the object-oriented features provided by C++ were therefore not fully utilized. In addition, as the C++ language evolved, there was no effort to update the \mpi{} C++ bindings. As a result, the \mpi{} C++ bindings were not widely used by newly designed C++ applications.  \mpi{} implementers were faced with a daunting task of supporting these bindings that were not widely used. Eventually, the \mpi{} Forum decided to deprecate the \mpi{} C++ bindings in the \mpi{}-2.2 \cite{MPI-2.2} standard and removed this feature in the \mpi{}-3.0 \cite{MPI-3.0} standard.  
\section{The C Interface Holds C++ Programmers Back}\label{sec:badc}

Using \mpi{} in a C application is not an issue. But, 
 many issues arise for C++ programmers using the C interface:
\begin{itemize}
\item {All C++ data structures need either to  be converted to or be native to C}.

\item {One cannot use a huge part of the C++ language: templates, classes, type specialization, lambdas, polymorphism, exceptions, iterators, ranges, etc.}

\item Info arguments and assertions  that could be compile-time-optimized must be runtime analyzed with C. 





\item {Efficient C++ constructs, such as smart pointers and reference-counted arrays, are useless with the C interface. All memory and scope management of objects and aliases are manual and must be handled with extreme care whereas these are easy to manage in modern languages.}

\item There is no encapsulation. MPI specifies a global set of functions, without  name-spacing; many  procedures  take weakly typed request objects, which are  polymorphic. 

\item  C is not flexible/expressive enough to address 64-bit support vs. 32-bit legacy issues. Big count needed to double up 157 C interfaces. C++ and modern  Fortran have overloading and largely skirt this issue.
\end{itemize}

It also should be noted the purely layered C++ interfaces aren't enough, because a native C++ library should be optimizable to be as fast or faster than a C interface when coupled with a suitable MPI implementation\footnote{Such performance-gain exploration is underway with ExaMPI \cite{ExaMPI}, which is purely C++ down to device drivers.}. Therefore, a standard language interface is needed to justify optimizing the critical path for C++ programmers.


\section{Background and Related Work}\label{sec:related}
C++ bindings for MPI began while MPI-1 was still being ratified
\cite{Bangalore94mpi++:issues,Skjellum96explicitparallel,oompi,DBLP:journals/concurrency/SkjellumWLWBLSM01} and continued until the official introduction of the C++ bindings in MPI-2. After MPI-3 discarded the interface, work evidently resumed.
Prior and on-going efforts are noted that sought to provide third-party MPI  language interfaces  including Java \cite{DBLP:journals/concurrency/CarpenterGJSF00}, C++ (e.g., \cite{Bangalore94mpi++:issues,oompi}), Spark \cite{MPIgnite}, and Python \cite{MPI4py}.

\subsection{MPJ: MPI-like Message Passing for Java}
In Carpenter et al.\@ \cite{DBLP:journals/concurrency/CarpenterGJSF00}, the MPJ notation for Java was introduced, providing a means to program Java with MPI-1-type operations.  This project has  been inactive since 2000.

\subsection{Boost MPI}
Boost MPI is a historical, header-only library on top of \mpi{}-2 that leverages the Boost library~\cite{gregor2017boost}. The goal was evidently to provide a higher-level abstraction like C++ STL. Boost MPI has not been widely used by major \mpi{} applications due to its complex dependencies and use of serialization to handle user defined datatypes. Recent minor updates in 2019 follow minor updates in 2013, and end of active development in 2008. 

\subsection{MPP}
MPP~\cite{MPP} is a header only, C++ \mpi{} interface that uses some of the object oriented programming features of C++ such as generic programming, type traits, futures, and also supports the use of user defined datatypes. Initial performance evaluation indicated better performance when compared with Boost MPI. This interface has not been updated since 2013. 

\subsection{MPL}
MPL~\cite{mpl} is an open-source, header-only, C++11-based implementation of a layered library on top of significant portions of  \mpi{}-3.1\footnote{Recent check-ins only as of May, 2021 have required C++17, but for superficial reasons such as adding \texttt{[[nodiscard]]} specifiers for returned values in certain functions.}.  The key features are as follows:
\begin{itemize}
    \item use of \texttt{comm} and \texttt{group} objects with \mpi{} APIs refactored into member functions (like \mpi{}-2 C++ interface)
    \item reduced argument sets compared to the C interface, including polymorphic variations with functions like gatherv (e.g., non-root processes specify less arguments)
    \item return of values and request objects in non-blocking operations, not error codes like C.
\end{itemize}
MPL does not support  one-sided, dynamic process management, or \mpi{} I/O.  Exceptions are thrown for various detected errors in MPL but there is no formal connection of the five exceptions it can throw to all the MPI error codes and classes.

Lastly, MPL provides useful abstractions built on top of \mpi{}, (e.g., logical process grids and a MPI derived datatype encapsulation).

\subsection{Ad Hoc C++ APIs}
A large-scale study of MPI usage in open-source HPC applications~\cite{10.1145/3295500.3356176} found that C++ was the most widely used language among open-source HPC applications. It appears that layering on top of the C interface is done commonly in C++-based MPI applications, particularly after MPI-3 deprecated the old C++ interface. For example, Comb~\cite{COMB} and HemeLB~\cite{HemeLB} are two particular examples of this trend.

\subsection{MPI4Py}
MPI4Py~\cite{MPI4py} provides a Python-friendly layer on top of \mpi{}-3.1, and works closely with NumPy~\cite{NumPy}. 
While MPI4Py enables access to \mpi{} for Python developers, it is heavily influenced by the C interface. However, the implementation of the futures package in MPI4Py shows an attempt at creating a more pythonic interface.
\section{A Cautionary Tale About Languages}\label{sec:consequences}

The introduction of new \mpi{} language interfaces  is an enticing opportunity for change. The benefits brought by this change can be of great benefit to those  currently tied to using a C interface from a language that is dissimilar to C, which forces design decisions and unidiomatic language usage.

This section aims to serve as a 
note of caution.
While introducing new features is required to advance any product, introducing too many leads to feature bloat, which which is something \mpi{} standardizers and users are intimately familiar. Therefore, the introduction of more languages and especially expressions of underlying concepts in these new languages that are unlike prior expressions must be thoroughly considered. The previously introduced C++ interface, which  was deprecated and removed one major version later, shows the danger.

The central problem to introducing more languages is that the current \standard{} has only  considered \mpi{} through the lens of the intersection of both C and Fortran. This has resulted in a document that fundamentally ties and limits the concepts seriously considered to what is expressible in C and Fortran. 

In fact, Fortran is held back in many ways given the languages advancement compared to C. If one were to decouple those two languages, accepting the cost of such an action (more complex language interaction), then each language could serve as its own interpretation of the concepts of \mpi{}.

To introduce any additional language or extend and change the current languages in an idiomatic way, we first need to conceptually separate the core \mpi{} from the expressions in C and Fortran. This is a formidable challenge in itself.

With the large variance in programming languages for which community language bindings have been developed, it is also important to consider that not all languages will support the same functionality. An example of this would be the profiling interface (PMPI); it provides an intercept functionality that is baked into C from the underlying method by which computers operate. However, in higher level languages such as Python, which is executed in the Python Virtual Machine, the functionality of PMPI can only be emulated and not replicated with decorators.


The MPI Forum must be cautious with introducing new language interfaces. The standard should not convert from its language-independent binding and C, Fortran bindings, to a C++-centric or Python-centric standard, per se. The key semantics and services of MPI should be contained in the core standard. The language interfaces should deliver these services in idiomatic ways without creating huge impacts on the main standard. In fact, with a division of the language-neutral specification from the language interfaces, the main document should shrink considerably in length and complexity.
\section{What does an idiomatic C++ language give us?}
\label{sec:idiomatic}


The C++ language has many powerful constructs and concepts that \mpi{} could take advantage of,
and such ideas are potential avenues for more performant or productive code. 
These ideas, some of which stray far from the current C bindings, are likely not to work with existing C++ applications using the C interface without code adaptations from \mpi{} libraries and applications; see Section~\ref{sec:mixing} for further discussion. 
Despite that caveat, some of these benefits include:
\begin{itemize}
    \item Pervasive Polymorphism---This enables extension libraries to provide the same \mpi{} object with new features
    \item Compile-time Optimizations---In a C++ interface, certain \texttt{MPI\_Info} objects could be treated as templates, and affect code paths at compile time instead of runtime. Additionally, \texttt{MPI\_Datatypes} could also have compile-time components. An example of this is \texttt{constexpr} plus C++20 ranges to create compile-time derived datatype specifications for faster runtime gather or scatter of non-contiguous data.
    \item Factory-based \mpi{} functions---  Certain \mpi{} functions could be altered to become factories for certain objects (such as for \texttt{MPI\_Requests}). For instance, all Sends and Receives could be generated by a general factory for that kind of operation; even blocking ones could fit this model with some special thinking for that case.
    \item Delegates---We could endow each object a set of allowable actions that return other objects with other/different allowable actions. This would help both \mpi{} application developers and compilers know at compile time what actions can and cannot be performed compliant with the standard. Concepts, introduced in C++20, may also be useful in this regard.
\end{itemize}

\section{zeroth interface in C++}\label{sec:zeroth}
%


The first incremental change from a C interface to a C++ interface introduces overloaded \mpi{} procedures with a ``Big Count'' variant. An example is shown below in Listing~\ref{lst:big_count}. Here, the procedure names remain the same while the definitions change. The specific procedure will be automatically picked by the compiler based on the variables passed by the user. With that change, most C++ applications could easily stop using the \texttt{\_c} procedures and maintain support for large counts, or gain support for them if they weren't using them.

\begin{lstlisting}[style=DerekStyle, label={lst:big_count}, caption={Example of overloaded C++ \mpi{} bindings}]
/* MPI 4.0 C Bindings */
int MPI_Type_size(MPI_Datatype datatype, int *size)
int MPI_Type_size_c(MPI_Datatype datatype, MPI_Count *size)
/* Potential C++ Bindings */
int MPI_Type_size(MPI_Datatype datatype, int *size)
int MPI_Type_size(MPI_Datatype datatype, MPI_Count *size)
\end{lstlisting}

The remaining examples below are more design-specific decisions on how to implement \mpi{} operations in a manner that is consistent with modern C++ style. As such, these examples are listed just to show possibilities, rather than purporting to be mature proposals for future C++ \mpi{} bindings.

The first such possible change the C++ interface would bring relates to how \mpi{} returns errors. In Listing~\ref{lst:big_count}, the function signatures still return an error code akin to the normal C bindings.  In C++, such errors could potentially be returned through exceptions (with the specific error code accessible in the caught exception). Exceptions would allow applications that don't care about errors to continue writing \mpi{} applications without change. For applications that are interested, they would use C++'s \texttt{try ... catch} mechanism instead of checking the return code manually. An example of this is shown in Listing~\ref{lst:errors}. Further, applications could also do multiple \mpi{} function calls inside a single try-block if they are willing to ignore  which of the enclosed \mpi{} procedures failed\footnote{It is possible to have the exception handler return the request of a failed operation. In that case, blocking requests would only be made programmer accessible when exceptions occur.}. Compatibility of this change with a C-based library is also possible. In the C interface, the implementation should capture the exceptions in the bindings, and return the error codes for the user; in short, the C binding should contain the \texttt{try ... catch} shown below.

\begin{lstlisting}[style=DerekStyle, label={lst:errors}, caption={Example of \mpi{} error codes in C++ }]
/* Current C way to check errors*/
int rc = MPI_Func();
if(rc != MPI_SUCCESS)
    MPI_Abort(MPI_COMM_WORLD, rc);
/* Potential C++ way to check errors */
try {
    MPI_Func();
} catch(MPIException e){
    int rc = /* from e */;
    MPI_Abort(MPI_COMM_WORLD, rc);
}
\end{lstlisting}

Next, if a C++ interface removes the obligation to return error codes, then that frees up some \mpi{} procedures to return values instead of requiring them to be \emph{OUT} parameters. For example, \texttt{MPI\_COMM\_RANK} would return an integer representing a process' rank inside a communicator, and \texttt{MPI\_COMM\_DUP} would return the new, duplicated communicator. These examples can be seen in Listing~\ref{lst:return_things}.

\begin{lstlisting}[style=DerekStyle, label={lst:return_things}, caption={Example \mpi{} functions returning objects and values in C++ }]
/* Possible function signatures */
int MPI_Comm_rank(MPI_Comm comm)
MPI_Comm MPI_Comm_dup(MPI_Comm comm)
/* Example usage of functions */
MPI_Comm dp = MPI_Comm_dup(MPI_COMM_WORLD);
int rank = MPI_Comm_rank(dp);
\end{lstlisting}
\section{Mixing C and C++ MPI}\label{sec:mixing}
%

There are, loosely speaking, two categories of \mpi{} C++ application developers.
The first consists of   programmers who choose to develop their entire application using modern C++ features and libraries. The second consists of those who aim to modernize or convert their application to C++, but must rely on a C-based library or libraries over which they have no control. This second category is the focus of this section. 

Currently, the  \mpi{} C bindings can be called from C++ applications without any complications. In the previous section, we introduced a set of changes that progressively utilize more C++ features.
Besides a few changes to specific functions, additional benefits of a C++ interface become limited by the second category of developers noted above if the C++ \mpi{} implementation must keep some of the C interface rules.
However, a C++ interface could provide both the original C interface and the new C++ interface by providing additional namespaces or headers that a library or application could use\footnote{A C++ implementation could also provide several versions of the C++ bindings.}.
An additional namespace requires that legacy C-based \mpi{} libraries or applications add a small wrapper that includes the code (e.g., \texttt{using namespace MPI\_C;}), while a different header requires a simple \texttt{\#include} change. 

The separate-header solution may indeed work for some applications. Some applications may wish to use the C++ bindings, but are stuck with a library that uses the C bindings and requires base \mpi{} objects to be passed to them (such as an \texttt{MPI\_Comm}). Such an application could also benefit from this dual-binding approach, but will require C++-to-C conversion functions for \mpi{} objects to be provided by the implementation. To support such applications, the C++ \mpi{} API would require functions to convert  objects from their C++ representation to their C (or Fortran) representation (or vice versa\footnote{Fortran to C++ conversions may not be required.}). A separate namespace for the bindings would also help limit any potential mixing of \mpi{} objects from different bindings without conversion. For example, the underlying type implementing  an~\texttt{MPI\_Comm} in the C bindings namespace would be incompatible in a function argument that takes a~\texttt{MPI\_Comm} as defined in the C++ bindings.

Lastly, we recommend  that, as \mpi{} supports more languages, such conversion functions be kept to a minimum to avoid namespace pollution in \mpi{}. Since C++ is still close to C, such conversion functions could prove useful to  developers, but conversion functions to/from other languages may not be as straightforward or needed as often as between C and C++. 
\section{Conclusions}
\label{sec:conclusions}
This paper motivated the need for a first-class C++20-based language interface for \mpi{}, and strong consideration of such interfaces for other modern languages.  Historical and ad hoc systems were discussed and possible design options for C++ in particular were mentioned. The next major standard release for MPI should seriously consider  modern language interfaces for multiple languages, and permit these to decouple from the extant language-independent specification that is firmly rooted in C89 and Fortran-77, and those languages' idiosyncrasies.

\section*{Acknowledgment}
\small
This work was performed with partial support from the National Science Foundation under Grants Nos.~CCF-1562306, CCF-1822191, CCF-1821431, and OAC-1925603. Any opinions, findings, and conclusions or recommendations expressed in this material are those of the authors and do not necessarily reflect the views of the National Science Foundation. 



%

\newpage

\bibliographystyle{IEEEtran}
\bibliography{references}

\end{document}